\documentclass[prb,twocolumn,showpacs,floatfix]{revtex4}
\usepackage{graphicx}
\usepackage{amssymb}
\usepackage{amsmath}
\usepackage{xspace}
\usepackage{url}

\newcommand{\ii}{\mathrm{i}}

\begin{document} 
\title{One-Dimensional Quantum Transport Affected by a Background Medium: Fluctuations versus Correlations}
\author{S.\ Ejima}
\affiliation{Institut f{\"ur} Physik,
             Ernst-Moritz-Arndt-Universit{\"a}t Greifswald,
             17489 Greifswald,
             Germany}
\author{H.\ Fehske}
\affiliation{Institut f{\"ur} Physik,
             Ernst-Moritz-Arndt-Universit{\"a}t Greifswald,
             17489 Greifswald,
             Germany}

\date{\today}
\begin{abstract}
We analyze the spectral properties of a very general two-channel 
fermion-boson transport model in the insulating and metallic regimes, 
and the signatures of the metal-insulator quantum phase transition 
in between. To this end we determine the single particle spectral
function related to angle-resolved photoemission spectroscopy, 
the momentum distribution function, 
the Drude weight and the optical response by means of a dynamical 
(pseudo-site) density-matrix renormalization group technique for 
the one-dimensional half-filled band case. We show how the interplay 
of correlations and fluctuations in the background medium controls
the charge dynamics of the system, which is a fundamental problem  
in a great variety of advanced materials.   
\end{abstract}
\maketitle
\section{Introduction}
Charge transport normally takes place in some background medium.
To understand how the environment affects the moving carrier
and vice versa is a difficult question and in this generality 
at present perhaps one of the most heavily debated issues in 
condensed matter physics. Here the term ``background''    
describes a variety of situations. We can think of the 
motion of a hole through an ordered insulator~\cite{Be09}.
Examples are the high-$T_c$ cuprates and 
the colossal magnetoresistive manganates, with a background 
of spins and orbitals, respectively, forming a pattern of
alternating order. Then, as the hole moves, it disrupts the order
of the background, which on its part hinders the particle 
transfer. Nevertheless coherent particle transport may 
occur, but on a strongly renormalized energy scale.   
The new quasiparticles formed in the cuprates and manganates 
are spin or orbital polarons~\cite{KLR89,MH91a,WOH09}. Another situation 
concerns a charge carrier coupled to a deformable background. 
Here, if the interaction with 
phonons is strong, the particle 
has to carry a phonon cloud through the medium. The outcome
might be a ``self-trapped'' small lattice polaron~\cite{Alex07}. 
In this case hopping transport, accompanied by phonon emission 
and absorption processes, evolves as the dominant transport channel.

So far we have considered a single particle only. It is quite 
obvious that the problem becomes even more involved if the particle
density increases. Then the inter-relation between charge carriers 
and background medium may drive quantum phase transitions. 
The appearance of ferromagnetism in the three-dimensional (3D) manganates,
superconductivity in the quasi-2D cuprates, or charge-density-wave
(CDW) states in 1D halogen-bridged transition metal complexes 
are prominent examples~\cite{TNFS00}. In the theoretical description
of these strongly correlated systems an additional difficulty arises: 
The particles which are responsible for charge transport and the 
order phenomena of the background are the same. As a consequence,
on a microscopic level, rather involved many-particle models  
result, which incorporate the coupling between charge, spin, orbital 
and lattice degrees of freedom~\cite{WOH09,WF04b}. 
Naturally this prevents an exact solution of the problem 
even in reduced dimensions.
\section{Model and Method}
\begin{figure}[b]
\includegraphics[width=0.85\linewidth]{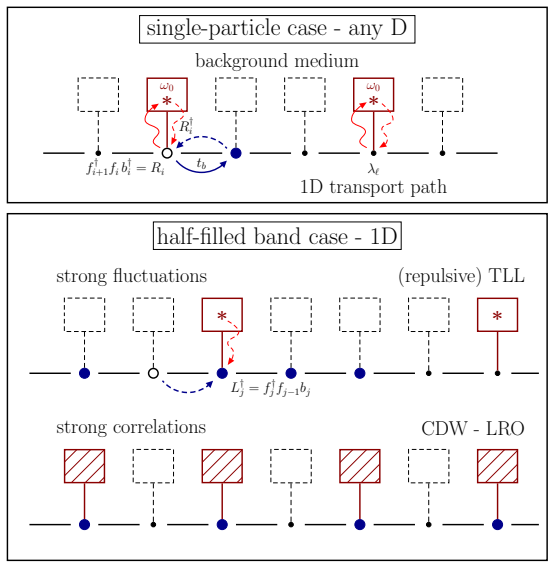}
\caption{(Color online) 
Schematic diagram of quantum transport in a background medium.
The background could represent a magnetically, orbitally or charge 
ordered lattice, but also a heat bath or certain chemical side groups. 
Then the proposed transport model~(\ref{model}) describes a very 
general situation: As a charge carrier ($\bullet$) moves 
along a 1D transport path it creates an excitation with energy $\omega_0$ 
($\ast$) in the background 
medium at the site it leaves or annihilates an 
existing excitation at the site it enters. 
It is a plausible assumption that the (de)excitation of the background 
can be parameterized as a bosonic degree of freedom  
. In the case of spin deviations, orbital
fluctuations or lattice vibrations, the bosons might be viewed as a 
Schwinger-bosons, orbitons or phonons. 
Of course, any distortion of the background can heal out 
by quantum fluctuations. Accordingly the $\lambda$-term 
allows for spontaneous boson creation and 
annihilation processes. The upper panel displays 
the single-particle case. Depending on the model
parameters quasi-free, diffusive or boson-assisted transport
takes place~\cite{AEF07}. The latter case corresponds e.g.  
to the motion of a hole through an ordered antiferromagnetic 
insulator. The lower panel shows the half-filled 
band case. Here, for spinless fermions in 1D, a repulsive 
Tomonaga-Luttinger liquid evolves, provided the excitations of the background 
are energetically inexpensive ($\omega_0 < \omega_{0,c}$) 
or will readily relax ($\lambda > \lambda_c(\omega_0)$).
This defines the fluctuation dominated regime.  
By contrast, strong background correlations, which develop for 
large $\omega_0$ and small $\lambda\ll t_b$ tend to 
immobilize the charge carriers and even may drive 
a metal insulator transition by establishing CDW 
long-range order~\cite{WFAE08,EHF09}.
\label{fig:model}}
\end{figure}

A way out might be the construction of simplified
transport models, which capture the basic mechanisms of 
quantum transport in a background medium in an effective way. 
Along this line a novel quantum transport model has been 
proposed recently~\cite{Ed06,AEF07}, 
\begin{equation}
 H= \!-t_b\sum_{\langle i, j \rangle} f_j^{\dagger}f_{i}^{}
  (b_i^{\dagger}+b_j^{})-\!\lambda\sum_i(b_i^{\dagger}+b_i^{})
              +\!\omega_0\sum_i b_i^{\dagger}b_i^{}\,,
\label{model}
\end{equation} 
which mimics the correlations inherent to a spinful fermionic many-particle 
systems 
by a boson affected hopping of spinless particles 
$\propto t_b$ (see Fig.~\ref{fig:model}). 
In the model~(\ref{model}), a fermion 
$f_i^{(\dagger)}$ creates (or absorbs) a local boson 
$b_i^{(\dagger)}$ every time it hops, which corresponds 
to a local excitation in the background with a certain energy 
$\omega_0$. Because of quantum fluctuations the distortions are 
able to relax $\propto \lambda$.  A unitary transformation
$b_i\to b_i + \lambda/\omega_0$ replaces this term by 
second transport channel $H_f=-t_f\sum_{\langle i, j \rangle} 
f_j^{\dagger}f_{i}^{}$, describing unaffected fermionic transfer, however    
with a renormalized amplitude $t_f=2\lambda t_b/\omega_0$. 
It has been shown~\cite{AEF07} that coherent propagation of a fermion 
is possible even in the limit $\lambda=t_f=0$
by means of a six-step vacuum-restoring hopping process,
\begin{equation} 
R_i^{(6)} = L^\dagger_{i+2} L^\dagger_{i+1} R^\dagger_i L_{i+2} R_{i+1} R_i\,,
\label{6-step}
\end{equation} 
where $R_i^{\dagger}=f_{i}^{\dagger}f_{i+1}^{}b_i^{}$ and 
$L_i^{\dagger}=f_{i}^{\dagger}f_{i-1}^{}b_i^{}$.
Note that $R_i^{(6)}$ acts as direct next neareast neighbor (NNN) 
transfer ``$f_{i+2}^{\dagger} f_i$'' , in complete analogy to the  
``Trugman path'' of a hole in a 2D N\'{e}el-ordered spin 
background~\cite{Tr88}. 

The model~(\ref{model}) has been solved 
in the single particle sector ($N_e=1$) by exact diagonalization~\cite{AEF07}, 
using a basis construction for the fermion-boson (many-particle)
Hilbert space that is variational for an infinite 
lattice ($N=\infty$)~\cite{BTB99}. The transport behavior was found to be
surprisingly complex, reflecting the properties of both 
spin and lattice polarons in $t$-$J$- and Holstein-type models.

For the 1D half-filled band sector ($N_e=N/2$), evidence for a metal insulator
transition comes from small cluster diagonalizations~\cite{WFAE08}.
Quite recently the ground-state phase diagram of the model~(\ref{model}) 
has been mapped out in the whole $\lambda$-$\omega_0$ 
plane~\cite{EHF09}, using a density-matrix renormalization group (DMRG) 
technique~\cite{Wh92}. A quantum phase 
transition between a Tomonaga-Luttinger liquid (TLL) 
and CDW was proven to exist. A complementary study of 
the dynamical properties of the system is therefore desirable. 

In the present work, we employ the dynamical DMRG (DDMRG) 
method~\cite{Je02} 
in order to investigate the effects of background fluctuations 
and correlations on the dynamics of charge carriers in the framework
of the  1D half-filled fermion-boson model~(\ref{model}).  Thereby
the focus is on the wave-vector resolves single-particle spectral
function probed by angle-resolved photoemission spectroscopy (ARPES) 
and on the optical conductivity probed e.g. by reflectivity measurements. 

In general the dynamic response of a quantum system described by 
a time-independent Hamiltonian $H$ is given by the imaginary part 
of correlation functions of type  
\begin{equation} 
 A_O(\omega)=\lim_{\eta\to 0}\frac{1}{\pi}
              \langle \psi_0 | O^{\dagger}
	       \frac{\eta}{(E_0+\omega-H)^2 +\eta^2}
	      O |\psi_0\rangle\, ,
\label{dynresp}
\end{equation}
where the operator $O$ identifies the physical quantity
of interest. $|\psi_0\rangle$ and $E_0$ give the 
ground-state wave function and energy of $H$. The small 
$\eta >0$ shifts the poles of the related Green function 
$G_O(\omega+\ii \eta)$ into the complex plane.

Single particle excitations associated with the injection or emission 
of an electron with wave vector $k$, $A^{+}(k,\omega)$ 
or $A^{-}(k,\omega)$, can be written in the spectral form  
\begin{equation}
 A^{\pm}(k,\omega)
=\sum_n|\langle\psi_n^{\pm}|f^{\pm}_k|\psi_0\rangle|^2\,
\delta[\omega\mp\omega^{\pm}]\,,
\label{spsfpm}
\end{equation}
where $f^+_k=f^\dagger_k$, $f^-_k=f^{}_k$.  $|\psi_0\rangle$ 
is the groundstate of a $N$-site system in the $N_e$-particle sector 
while $| \psi_n^{\pm}\rangle$ denote the $n$-th 
excited states in the $N_e\pm 1$-particle sectors 
with excitation energies $\omega_n^\pm=E_n^\pm-E_0$.

Optical excitations, on the other hand, connect states
in the same particle sector with a site-parity change.
For a system with open boundary conditions (OBC) the regular part of 
the optical absorption
\begin{equation}
 \sigma_{reg}(\omega)=\frac{\pi}{N}\sum_n \omega_n\,
                           |\langle\psi_n|P|\psi_0\rangle|^2\,
                           \delta[\omega-\omega_n]
\label{sigmareg1}
\end{equation}
is related to the dynamical polarizability, $\sigma_{reg}(\omega)=\omega \alpha(\omega)$, where $P=-\sum_{j=1}^N j\,(f_j^\dagger f_j^{} -1)$ is the dipole 
operator (in units of $e$), and $\omega_n=(E_n-E_0)$. Then 
the current operator is obtained from $J= \ii [H,P]$. 
Applying periodic boundary conditions (PBC), the optical conductivity
can be calculated from
\begin{equation}
 \sigma_{reg}(\omega)=\frac{\pi}{N}\sum_n 
                           \frac{|\langle\psi_n|J|\psi_0\rangle|^2}{\omega_n}\,
                           \delta[\omega-\omega_n].
\label{sigmareg2}
\end{equation}
Note that for our fermion-boson model~(\ref{model}), the
current operator has two contributions,  $J=J_f+J_b$, 
where $J_f=\ii t_f\sum_j f_{j+1}^{\dagger}f_j^{\phantom{\dagger}}
      -f_{j}^{\dagger}f_{j+1}^{\phantom{\dagger}}$ and $J_b=\ii t_b\sum_j 
       f_{j+1}^{\dagger}f_j^{\phantom{\dagger}}b_{j}^{\dagger}
      -f_{j}^{\dagger}f_{j+1}^{\phantom{\dagger}}b_{j}^{\phantom{\dagger}}
      +f_{j-1}^{\dagger}f_{j}^{\phantom{\dagger}}b_{j}^{\dagger}
      -f_{j}^{\dagger}f_{j-1}^{\phantom{\dagger}}b_{j}^{\phantom{\dagger}}$.
The f-sum rule 
\begin{equation}\label{EqFSum}
   S_{reg}(\infty)  + \pi D
= -\pi E_{kin}/2 
\end{equation}
connects the frequency-integrated optical response
$S_{reg}(\omega)=\int_{0}^\omega \sigma_{reg}(\omega^\prime) \,d\omega^\prime$
to the kinetic energy 
$E_{kin} = \frac{1}{N}\langle 0|H-\omega_0 \sum_i b^\dagger_i
b^{}_i|0\rangle$, where the Drude part $\propto D$ 
serves as a measure for coherent transport. For OBC,
only a $D$ precursor exists in the metallic region. 

In the actual DDMRG calculation of spectral functions
the required CPU time increases rapidly with the number 
of the density-matrix eigenstates $m$. 
Since the DDMRG approach is based on a variational 
principle~\cite{Je02}, we first of all have to prepare  
a good ``trial function'' for the ground state with as 
many density-matrix eigenstates as possible. 
As a rule we keep $m\sim 500$ states to obtain the true ground state
in the first five DDMRG sweeps and afterwards take $m \sim 200$ states
for the calculation of the various spectra from~(\ref{dynresp}) with 
a broadening $\eta=0.1$. In order to save CPU time in the DDMRG runs 
we take into account just $n_b=3$ pseudosites.
In this case the $n_b$-th local boson pseudo-site density is smaller 
than $10^{-5}$. Using $n_b=4$ this value can be reduced to $10^{-8}$ 
which leads, however, not to visible change of the spectra because the 
discarded weight in the DDMRG calculations is $\sim 10^{-3}$
(i.e. three orders of magnitude larger than for the DMRG 
ground-state calculations).
 
\section{Results}
\subsection{Photoemission spectrum}
Let us first discuss the single-particle spectra
of the transport model~(\ref{model}) in the regime where the 
background is stiff, i.e. the distortions 
induced by the particle hopping process 
are energetically costly ($\omega_0=2$). 

\begin{figure}[t]
\includegraphics[width=0.98\linewidth]{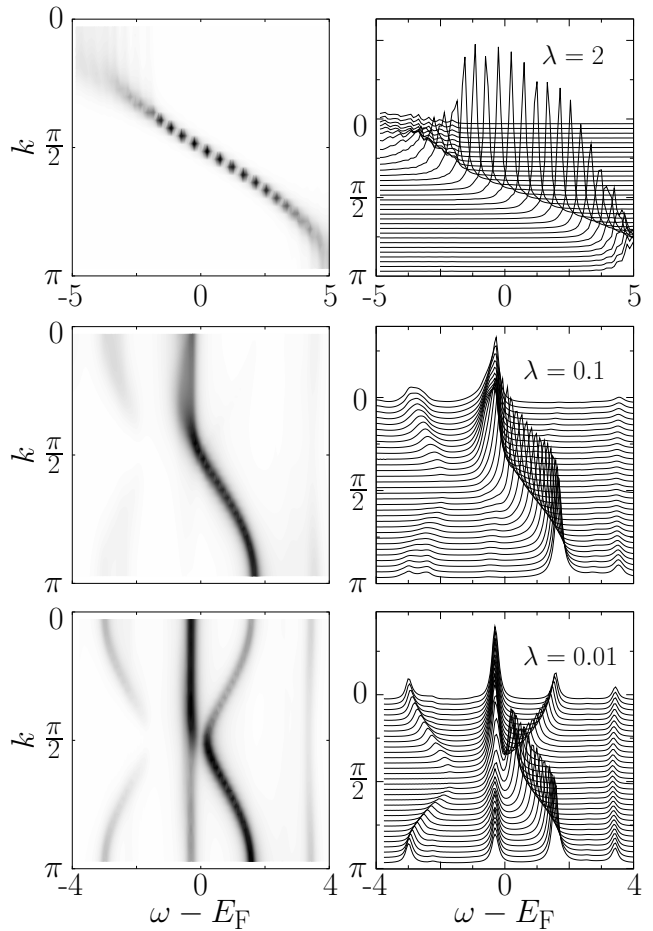}
\caption{Intensity (left panels) and line-shape (right panels) 
of the single-particle spectral function $A(k,\omega)$
in the half-filled band sector of the fermion-boson transport
model~(\ref{model}) on a $N=32$--site chain. 
The upper two rows (lower row) give DDMRG data  
for  $\lambda=2$ and $0.1$ in the metallic regime 
($\lambda=0.01$, insulating regime), where $\omega_0=2$.
All energies are measured in units of $t_b$.
Since we apply OBC, we use quasi-momenta $k=\pi l / (N+1)$
with integers $1\leq l\leq N$. 
}
\label{fig:Akw-w2.0}
\end{figure}

For very large $\lambda$ the free transport channel nevertheless dominates 
and an almost particle-hole symmetric spectrum 
($A^+(k,\omega-E_F)\sim A^-(k-\pi, E_F-\omega)$) 
results (see Fig.~\ref{fig:Akw-w2.0} upper panels). 
As $\lambda$ decreases, the background distortions hardly relax. 
Consequently the bosonic degrees of freedom will strongly affect the transport. 
The middle panels of Fig.~\ref{fig:Akw-w2.0} show how, at $\lambda=0.1$,
strong correlations develop in the occupied states probed by 
photoemission (PE) for $\omega < E_F$. The introduced hole  
can only move coherently by the 6-step process~(\ref{6-step}),
where in step 1-3 three bosons were excited, which are consumed in
steps 4-6 afterwards. In this way the collective particle-boson dynamics 
leads to a flattening of the ``coherent'' band for $k\leq k_F$. 
By contrast an additional electron, which probes the unoccupied
states in an inverse (I)PE experiment ($\omega > E_F$), 
can more easily move by a two-step process, even if pronounced 
CDW correlations exist in the background medium~\cite{WFAE08}. 
The incoherent parts of  $A(k,\omega)$ far away from the Fermi energy 
$E_F$ are caused by excitations with additional bosons 
involved (bear in mind that the ground state with $N_e$ electrons 
is a multi-phonon state and the wave vector of the $N_e\pm 1$ 
target state corresponds to the total momentum of electrons and bosons). 

While for $\lambda=0.1$, $A(k_F,\omega)$ has finite spectral weight 
at $E_F$, i.e. the system is still metallic (albeit the
TLL charge exponent $K_{\rho}$ is noticeably reduced from one~\cite{EHF09}),  
an excitation gap opens in the PE spectrum as 
$\lambda$ falls below a certain critical value, provided 
that $\omega_0 > \omega_{0,c}(\lambda=0)$~\cite{EHF09}.
We find $\lambda_c(\omega_0=2)\simeq 0.05$.  
The lower panels of Fig.~\ref{fig:Akw-w2.0} show $A(k,\omega)$ 
for $\lambda=0.01$, in the insulating regime, where a 
CDW with true long-range order exists. The TLL--CDW 
quantum phase transition is driven by the correlations 
that might evolve in the background medium at commensurate 
fillings. Let us emphasize the dynamical aspect of this 
process: The (collective) bosonic excitations are intimately 
connected to the motion of the particles, and themselves
have to persist long enough in order to affect the many-particle state.
\begin{figure}[t]
\includegraphics[width=0.95\linewidth]{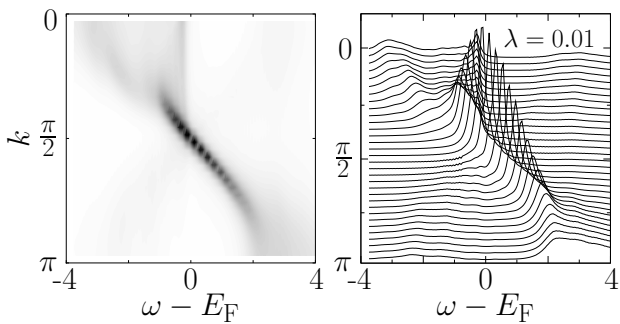}
\caption{Density (left) and line shape (right) plot
of the $A(k,\omega)$ spectra. Again $N=32$ (OBC), 
$\lambda=0.01$, but now $\omega_0=1$.}
\label{fig:Akw-w1.0}
\end{figure}
The ARPES spectrum for the insulating state clearly 
shows the doubling of the Brillouin zone. The remaining 
asymmetry with regard to the spectral weight of the absorption 
signals as $k\leftrightarrow (\pi-k)$ vanishes for $\lambda\to 0$. 
Most notably the widths of the highest PE and lowest IPE coherent 
bands differ by a factor of about $(t_b/\omega_0)^4$, since the 
CDW order is restored if the injected hole [electron] is transfered 
to a NNN site by a process of order $O(t_b^6/\omega_0^5)$ 
[$O(t_b^2/\omega_0)$]. Hence the CDW state exhibits a correlation-induced 
asymmetric band structure~\cite{WFAE08}.   

The strong interrelation of charge dynamics and background fluctuations
becomes obvious if we decrease $\omega_0$ below $\omega_{0,c}$  
keeping $\lambda=0.01$ fixed. Of course, in passing the accompanied 
insulator-metal transition the PE spectrum changes completely, 
but the ``nature'' of 
the TLL at $\omega_0=1$ is different compared to  that of the 
metallic state realized at larger $\omega_0$ and $\lambda$ as well  
(cf. Fig.~\ref{fig:Akw-w1.0} and upper panels of Fig.~\ref{fig:Akw-w2.0}). 
The single-particle spectrum for $\omega_0=1$ shows sharp absorption 
signals in the vicinity of $k_F$ only. In a wide $k$-space region 
emanating from $k=0$ ($k=\pi$) the PE (IPE) spectrum
is smeared out (over-damped), i.e. here the dynamics of 
the system is dominated by bosonic fluctuations. 
\subsection{Momentum distribution function}
\begin{figure}[t]
\includegraphics[width=0.98\linewidth]{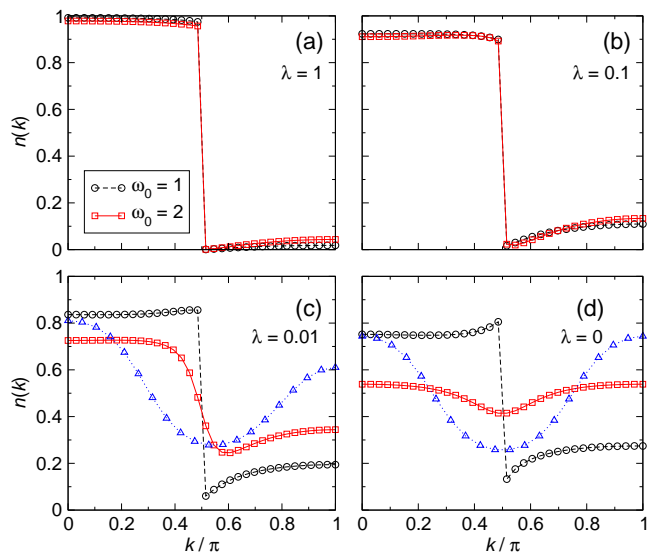}
\caption{(Color online) Momentum distribution function $n(k)$ 
for the half-filled two-channel transport model~(\ref{model})
with 66 sites and PBC, as $\lambda$ decreases 
from one~(a) to zero (d) at $\omega_0=1$ (circles) and~2 (squares).   
Triangles show $n(k)$ for a half-filled $t$-$t^\prime$-$U$ model  
(38 sites, PBC)  with $U=10t^\prime$, $t=t^\prime$ [panel (c)] 
and $t=0$ [panel (d)] (see text). In order to obtain more accurate 
ground-state data we use $n_b=4$ pseudo-sites.}
\label{fig:nk-Edwards}
\end{figure}

The different transport behavior becomes also apparent in the 
momentum distribution function  
\begin{equation}
n(k) =\frac{1}{N}\sum_{j,l}\mathrm{e}^{\ii k(j-l)}
        \langle 
	 c^\dagger_{j} c^{\phantom{\dagger}}_{l}
        \rangle\,.
\end{equation}
By means of DMRG, the ground-state correlation function
$\langle c^\dagger_{j} c^{\phantom{\dagger}}_{l}\rangle$  
can be easily calculated for PBC. Figure~\ref{fig:nk-Edwards}  
displays $n(k)$ for two characteristic boson energies, 
above and below $\omega_{0,c}$. 

In the former case, the 
TLL-CDW transition causes significant changes
in the functional form of $n(k)$. For $\lambda> \lambda_c$,
one expects an essential power law singularity at $k_F$, 
corresponding to a vanishing quasiparticle weight.
For finite TLL systems the difference $\Delta= n(k_F-\delta)
-n(k_F+\delta)$ is finite (with $\delta=\pi/66$ in our case)~\cite{EF09a}.
$\Delta$ rapidly decreases approaching the CDW transition point  
with decreasing $\lambda$ (see data for $\omega_0=2$ (red squares)). 
In the CDW phase the singularity at $k_F$ vanishes. Note that 
the periodicity of $n(k)$ doubles at $\lambda=0$,
in accordance with a $R^{(6)}$ NNN-only hopping channel.  
To substantiate this reasoning we have included in 
Fig.~\ref{fig:nk-Edwards}  $n(k)$--data calculated for the 1D 
Hubbard model with additional NNN-transfer $t^\prime$. 
We see that $n(k)$ of the fermion-boson model~(\ref{model})  
is in qualitative agreement with our data and previous 
results for the $t$-$t^\prime$ Hubbard model~\cite{GHW05},
in particular for the case $t=0$. The upturn in $n(k)$ for
$k>k_F$ persists even in the metallic regime as long as 
NNN hopping processes triggered by the (CDW) correlations
in the background are of importance. 

For $\omega_0=1$ the system stays metallic for all $\lambda$. 
Besides the usual renormalization of $n(k)$ with increasing 
correlations (i.e. decreasing $\lambda$) we find a slight 
upturn in $n(k)$ for 
$k \lesssim k_f$. This might be attributed to the fact
that in our model~(\ref{model}) a particle injected with 
$k=\pm \pi$ is almost unaffected by bosonic fluctuations
(which holds also for the single-particle case~\cite{AEF07}).
So to speak the system behaves as a nearly perfect metal at 
this point. It is worth to mention that
an increase in $n(k)$ for both $k\lesssim k_F$ and
$k\gtrsim k_F$ has been found also for the momentum distribution 
function of the Hubbard model (with and without magnetization) 
using the Gutzwiller variational wave function~\cite{MV87}.     
\begin{figure}[htbp]
\includegraphics[width=0.95\linewidth]{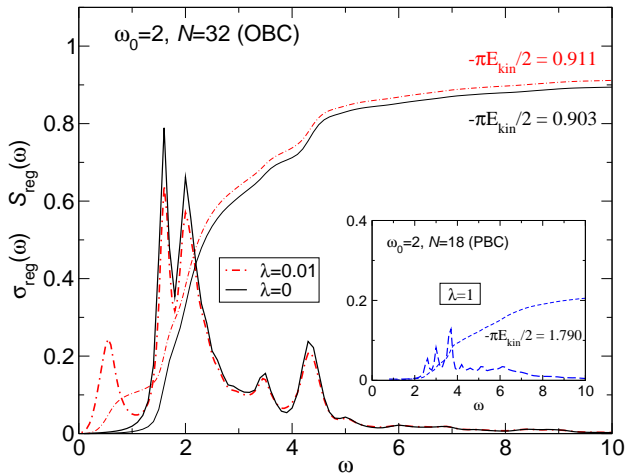}
\caption{(Color online) Optical conductivity $\sigma_{reg}(\omega)$
and integrated spectral weight  $S_{reg}(\omega)$
in the half-filled transport model~(\ref{model}).
The main panel (inset) displays DDMRG data for $\omega_0=2$
at $\lambda < \lambda_c$  ($\lambda \gg \lambda_c$), i.e. 
in the insulating (metallic) region, applying OBC (PBC).}
\label{fig:sigmaomega}
\end{figure}
\subsection{Optical response}
Finally we consider the evolution of the optical conductivity
going from the correlated TLL to the CDW phase at $\omega_0=2$. 
The corresponding optical absorption spectra are depicted
in Fig.~\ref{fig:sigmaomega}. In the metallic state most of 
the spectral weight resides in the coherent Drude part.
At $\lambda=1$ (see inset), we find $\pi D/N \simeq 1.6$, which has to 
be compared with $S_{reg}(\infty)\simeq 0.2$
(of course $D$ decreases as $\lambda$ gets smaller). 
In this case the wave-vector resolved single-particle spectra 
roughly extends from $\omega=-6$ to $\omega=6$. The regular part of the 
conductivity is mainly due to excitations to the phononic side bands
appearing in the sectors with momenta far away from $k_F$.     
In the insulating region (see main panel), the first
peak at about $\omega\simeq 0.5$ can be assigned to an 
optical excitation across the gap in the (coherent) two-band 
structure. These excitations are only 
accessible for $\lambda >0$. Additional excitations with higher energy 
occur around multiples of the boson frequency, where 
$\omega\simeq \omega_0=2$ sets a absorption threshold for the 
$\lambda=0$ case. As expected for an insulating system
with OBC, the whole spectral weight is contained in 
$S_{reg}(\infty)\simeq -\pi E_{kin}/2$. We emphasize that the 
CDW state in our model contains less than one boson per site
on average, unlike e.g. the Peierls insulating state
in the Holstein model. That is the CDW phase typifies rather as a  
correlated insulator---such as the Mott-Hubbard insulator---and 
no divergence occurs at the optical absorption threshold~\cite{JGE00}. 

\section{Summary}
In conclusion, we have determined the spectral properties
of a highly non-trivial two-channel fermion-boson transport
model for the 1D  half-filled band case, using an unbiased 
DDMRG technique.  The background medium, parameterized by bosonic 
degrees of freedom, strongly influences the charge carrier dynamics,
as it happens in many novel materials. If the background fluctuations 
dominate we find diffusive transport. In opposite case of strong
background correlations coherent quantum transport may evolve on 
a reduced energy scale. These correlations can also trigger 
a metal insulator transition. The insulating CDW state has a 
asymmetric band structure, leading to characteristic 
signatures in the ARPES and optical response. Whether an extended model 
with spinful fermions gives rise to an attractive metallic phase
like in the Holstein-Hubbard model~\cite{TAA05}  would be an 
interesting question for further research. 

\section*{Acknowledgments}
The authors would like to thank A. Alvermann,  K. W. Becker, D. M. Edwards, 
F. Gebhard, G. Hager, E. Jeckelmann, S. Sykora, L. Tincani,
S. A. Trugman and G. Wellein for valuable discussions. This work was
supported by DFG through SFB 652 and the KONWIHR project HQS@HPC.
\vspace*{-0.6cm}


\end{document}